# Thickness dependent magnetic anisotropy of ultrathin LCMO epitaxial thin films

Norbert M Nemes[1,2], Mar García-Hernández[2], Zsolt Szatmári[3], Titusz Fehér[3], Ferenc Simon[3], Cristina Visani[1], Vanessa Peña[1], Christian Miller[1], Javier García-Barriocanal[1], Flavio Bruno[1], Zouhair Sefrioui[1], Carlos Leon[1], Jacobo Santamaría[1]

[1]GFMC, Dpto. Fisica Aplicada III, Universidad Complutense de Madrid, 28040 Madrid, Spain
[2]Instituto de Ciencia de Materiales de Madrid, ICMM-CSIC, 28049 Cantoblanco, Spain
[3]Budapest University of Technology and Economics and Condensed Matter Physics Research Group of the Hungarian Academy of Sciences, P.O. Box 91, Budapest, H-1521

**The magnetic properties of $La_{0.7}Ca_{0.3}MnO_3$ (LCMO) manganite thin films were studied with magnetometry and ferromagnetic resonance as a function of film thickness. They maintain the colossal magnetoresistance behavior with a pronounced metal-insulator transition around 150-200 K, except for the very thinnest films studied (3 nm). Nevertheless, LCMO films as thin as 3 nm remain ferromagnetic, without a decrease in saturation magnetization, indicating an absence of dead-layers, although below approx. 6 nm the films remain insulating at low temperature. Magnetization hysteresis loops reveal that the magnetic easy axes lie in the plane of the film for thicknesses in the range of 4-15 nm. Ferromagnetic resonance studies confirm that the easy axes are in-plane, and find a biaxial symmetry in-plane with two, perpendicular easy axes. The directions of the easy axes with respect to the crystallographic directions of the cubic $SrTiO_3$ substrate differ by 45 degrees in 4 nm and 15 nm thick LCMO films.**

## I. INTRODUCTION

The growth of heterostructures combining oxide materials is a new strategy to design novel artificial multifunctional device concepts. In particular there has been growing interest in heterostructures involving oxide or elemental superconductors (S) and colossal magnetoresistance (CMR) ferromagnets (F) made of $YBa_2Cu_3O_7$ (YBCO) and $La_{0.7}Ca_{0.3}MnO_3$ (LCMO) [1-4]. The critical temperature of the superconductor depends on the relative orientation of the magnetization of the F layers, giving rise to a new giant magnetoresistance (GMR) effect which might be of interest for spintronic applications. The origin of this magnetoresistance is quite controversial and while some reports have proposed spin dependent effects on transport [3-6] others favor the importance on stray fields due to the domain structure of the ferromagnet in depressing the critical temperature at the coercive field [7-9]. A detailed study of the domain structure of the ferromagnetic layers is thus of major importance. Magnetic anisotropy may depend on crystalline orientation and on thickness through the influence of strain, surface morphology and roughness [10-13].

In this paper we focus on the magnetic properties of epitaxial LCMO thin-films as the first and necessary step in gaining an understanding of the properties of F/S structures. We address this with resistivity, magnetization and ferromagnetic resonance experiments (FMR).

## II. EXPERIMENTAL

Samples were grown on (100) oriented $SrTiO_3$ (STO) single crystals in a high pressure (3.4 mbar) dc sputtering apparatus at high growth temperature (900°C). The high oxygen pressure and the high deposition temperature provide a very slow (1 nm/min) and highly thermalized growth which allows control of the deposition rate down to the unit cell limit [14]. For this study, we grew LCMO single layers varying the thickness between 3 and 15 nm.

Structure was analyzed using x-ray diffraction and transmission electron microscopy. Further details about growth and structure can be found elsewhere [15,16]. An X-ray refinement technique, using the SUPREX 9.0 software, was used to obtain quantitative information about the interface roughness [17]. Thin manganite films on STO suffer compressive strain along the c-axis. This strain relaxes relatively slowly, the c lattice parameter changes only by 0.5% between a d=60A and a d=1000A LCMO film that in turn differs from the bulk by 1.2%. Preliminary AFM images (not shown) demonstrate that the surface morphology is smooth, following that of the STO substrate. TEM images of similar LCMO films show high quality, epitaxial structure [1,14-16]. Magnetization was measured in a VSM (Quantum Design PPPS) magnetometer. Temperature dependent magnetization was recorded on warming in a field of 0.1 T after field-cooling in 1 T. The hysteresis loops were recorded at 77 K and 10 K between either ±3T or ±1T. Magnetotransport was measured in a cryostat equipped with a 9T magnet (Quantum Design PPMS-9T). FMR was recorded in a modified JEOL ESR spectrometer operated at 8.9 GHz, in a cryostat filled with liquid nitrogen. The substrates were 10*5*1 mm$^3$ rectangles and were subsequently cut with a diamond wheel to smaller pieces as needed by the various techniques. X-ray scattering was measured on a 5*5 mm$^2$ piece and for resistivity measurements four silver contacts were evaporated in the corners. Magnetization experiments were done on 1.5*1.5 mm$^2$ squares to allow for measurements in orientation perpendicular to the magnetic field. For the FMR experiments, the 1.5*1.5 mm$^2$ pieces were mounted on small, cylindrical, teflon holders either parallel or perpendicular to the cylindrical axis using vacuum grease and then sealed in 4 mm diameter quartz tubes after evacuating and backfilling them with 20 mbar He exchange gas. These ampoules were then directly placed in the liquid nitrogen.

**FIG. 1 HERE**

### III. RESULTS

In Fig. 1 we summarize the magnetization (top) and resistivity (bottom) results of LCMO thin films of various thicknesses. Manganite layers as thin as 3 nm exhibit clear ferromagnetic hysteresis loops. The saturation magnetization of the thinnest manganite films takes values close to the bulk indicating the absence of thick magnetic dead layers. The coercive field increases drastically in thinner films, suggesting a change in the domain structure. Nevertheless, the field of saturation is also larger in thinner films raising the possibility that the magnetization lies out of the film plane. Temperature dependent resistivity is shown in the bottom panel of Fig. 1. Films thicker than 6 nm are metallic at low temperature with well defined metal insulator transition, although the thinnest films remain insulating in the whole temperature range.

**FIG. 2 HERE**

Next we investigate the direction of easy axes of magnetization in the individual F layers. Fig. 2 shows the magnetization hysteresis loops of 4 nm (top) and 15 nm (bottom) thick LCMO thin films. The hysteresis loops were recorded at 77 K, to match the temperature of the FMR experiments and also at 10 K (not shown). Each sample was mounted three times, on a standard Plexiglas VSM sample-holder using kapton-tape, with the external magnetic field directed along the [100], [110] and [001] crystallographic directions of the cubic STO substrate, with the first two lying in the plane of the LCMO thin film, and the third being perpendicular. The magnetic field was swept by 50 Oe/s. The data appear quite noisy, but this is due only to the very small size of the samples; the typical saturation magnetization is 1-2 *10$^{-5}$ emu, whereas the noise floor of the VSM is a few times 10$^{-7}$ emu. The hysteresis loops of both LCMO recorded with the field perpendicular to the film-plane exhibit a pronounced round shape and very large saturation field (~0.5 T and over 1 T for the 4 nm and the 15 nm thick LCMO, respectively), compared to the in-plane hysteresis loops (with saturation fields around 0.1 T). The coercive field is also appreciably larger for both samples with the field perpendicular. These observations indicate that the magnetization lies preferentially in-plane in both LCMO, irrespective of thickness [10-13]. Furthermore, similar differences can be discerned between the two in-plane hysteresis loops of the 15 nm thick LCMO (Fig. 2, lower panel). The lowest saturation field is seen when the magnetic field is oriented along a [110] direction of the cubic STO substrate, or practically, making 45 degrees with the edge of the sample. This indicates that the easy axes in these thicker LCMO films lie along the [110] axes. Similar conclusions can not be drawn for the in-plane anisotropy of the thinner LCMO based on the magnetization hysteresis loops because of the relatively stronger noise of the data. The difference between the in-plane and out-of-plane hysteresis loops is less pronounced in the 4 nm thick sample.

Fig. 3 shows a series of FMR spectra for both the 4 nm (left) and the 15 nm thick (right) LCMO thin films, as they are rotated in-plane with respect to the external magnetic field. 0, 90, 180 and 270 degrees indicate directions when the field is parallel to the side of the samples, and thus to crystal axes, [100] or [010], of the cubic STO substrate. Consequently, the field in these directions also coincides with pseudo-cubic crystal axes of the manganite. It is apparent from Fig. 3 that the position of the FMR varies with this in-plane angle. In fact, it displays 4 maxima and minima. Furthermore, at angles when the FMR is at maximum in the 4 nm thick LCMO, it is at minimum in the 15 nm thick LCMO and vice versa. The small peaks around 3300 G are paramagnetic resonance lines of the STO substrate. In Figs. 3 and 4 the y-axis position of the baseline denotes the angle of orientation in which the spectra were recorded, whereas the actual spectral intensity (the derivative of microwave absorption) is given in arbitrary units. The lower panel of Figure 3 shows the FMR field (indicated radially) extracted from the spectra on a polar plot, with the polar angle corresponding to the actual orientation of the samples in the spectrometer. Similarly, Fig. 4 depicts a series of FMR spectra as the film plane is rotated with respect to the external magnetic field. The lower panel shows the FMR field vs. out-of-plane orientation.

**FIG. 3 HERE**

Using ferromagnetic resonance we map the direction of the in-plane easy axes in 4 and 15 nm thick LCMO films and find that both exhibit biaxial symmetry, similar to La$_{0.7}$Sr$_{0.3}$MnO$_3$ (LSMO) thin films [10,11]. Nevertheless, the symmetry direction in the two films is rotated by 45 degrees with respect to the crystallographic directions. In the 4 nm film the in-plane easy axes are the [100] and [010] while in the 15 nm film the easy axes are the [110] and [1-10], indicated by the minimum values of the FMR field. We have been able to model the 15 nm data with a simple ferromagnetic Hamiltonian including a uniaxial anisotropy perpendicular to the film and a fourfold anisotropy in the plane of the film:

$H = -B_0 \cdot m - \tfrac{1}{2}\mu_0 M\, \check{n}\, N\, m + K_4(\check{n}_x^4 + \check{n}_y^4) + K_u\, \check{n}_z^2$.

Here $B_0$ is the external magnetic field, m is the magnetic moment of an Mn ion, ň is a unit vector parallel to m, M is the magnetization of the FM layer, N is the demagnetization tensor [$(N_{xx}, N_{yy}, N_{zz}) = (1,1,0)$ in the case of a film perpendicular to the z axis], and $K_4$ and $K_u$ are the fourfold and uniaxial anisotropies, respectively. For the curves in Figs. 3 and 4 we used M=0.45 A/m [M=450 emu/cm$^3$], $K_4$=0.0067 K and $K_u$=0.2 K [$K_u$=4.8*10$^5$ erg/cm$^3$]. Since demagnetization and uniaxial anisotropy give the same orientation dependence, we can also model the data by incorporating the shape anisotropy into the uniaxial anisotropy. Then $K_u$=0.39 K [$K_u$=8.8*10$^5$ erg/cm$^3$], similar to the results obtained in Ref. [13]. This model is unable to describe the FMR results of the 4 nm thick sample. The data in Fig. 3 deviate slightly from the 4-fold symmetry due to the inevitable misalignment of the film plane and the external magnetic field. As the sample is rotated, it may wobble to and from the field direction. The principal directions are also a few degrees off of the expected 0, 45, 90 etc. because of a similar uncertainty in the angular offset when the sample is mounted. Fig. 4 shows the position of the FMR as the external field is rotated out-of-plane and the corresponding model based on the above Hamiltonian.

**FIG. 4 HERE**

## IV. DISCUSSION

The demonstration of biaxial in-plane magnetic anisotropy in LCMO is itself a new contribution of the present study. It agrees well with the similar finding in LSMO thin films and with recent reports in LCMO thin films [13]. However, the rotation of the biaxial anisotropy axes by 45 degrees with the change of the thickness of the LCMO is a new and unexpected finding. The direction of the easy axes in the thicker (15 nm) LCMO film agrees with that found in LSMO, that is, they lie along axes such as [110] and [1-10]. There are several reasons why the easy axes move to the crystal axes [100] and [010]. Thinner LCMO is more strained. The STO substrate and bulk LCMO are not perfectly lattice matched ($a_{LCMO}$=3.86 A, $a_{STO}$=3.905A) and tensile epitaxial strain may influence the orbital structure (favoring the occupation of the $d_{x2-y2}$ orbitals) and thus influence the magnetic anisotropy. Strain relaxation in thicker samples may thus generate a change in the magnetic anisotropy. Nevertheless, it is also possible that the magnetic anisotropy of very thin LCMO layers is influenced by extrinsic mechanisms such as steps and other correlated defects of the substrate. The STO single crystal substrates are cut along the (100) crystal planes within an approx. 0.5° uncertainty. Therefore, lattice steps are present on the polished surface. However, the direction of these terraces is completely random, being determined by the exact way the substrate is mounted during the polishing step.

As the thickness of the manganite thin film is shown to be responsible for dramatic changes in the magnetic and metallic behavior of the ferromagnetic layer, it will be important to study the properties of manganite thin films when they form part of various LCMO/YBCO (F/S) hetero-structures.

In summary, ultrathin LCMO epitaxial films grown on STO remain ferromagnetic down to 3 nm. However, the metal-insulator transition is maintained only in films thicker than about 6 nm. The magnetic easy axis lies in the film plane independent of thickness, at liquid nitrogen temperature. However, there is a pronounced biaxial in-plane magnetic anisotropy, with two perpendicular, equivalent easy axes. The orientation of these easy axes changes with LCMO thickness. In thicker LCMO (~15 nm) the easy axes are the [110] and [1-10] pseudo-cubic directions. In thinner LCMO (~4 nm) the easy axes coincide with the in-plane crystal axes [100] and [010].

## ACKNOWLEDGMENT

N. N. M. acknowledges the "Ramon y Cajal" fellowship of the Spanish Ministry of Science and Education. This work was supported in part by SPIN-MAT06024 C01 and C02 and by Marie-Curie-IRG grant No. 024861, and by OTKA PF63954, K68807 and NK60984.

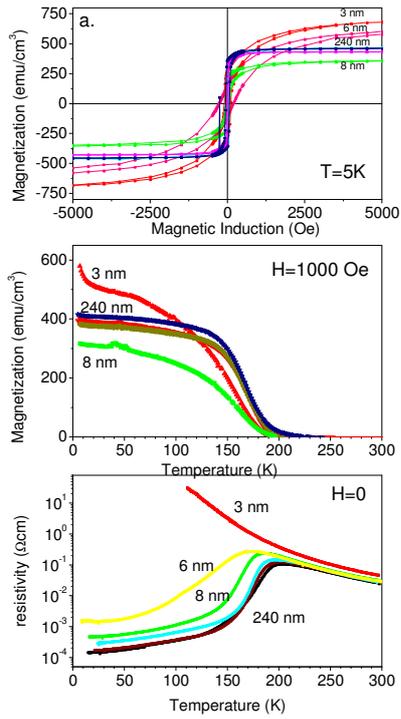

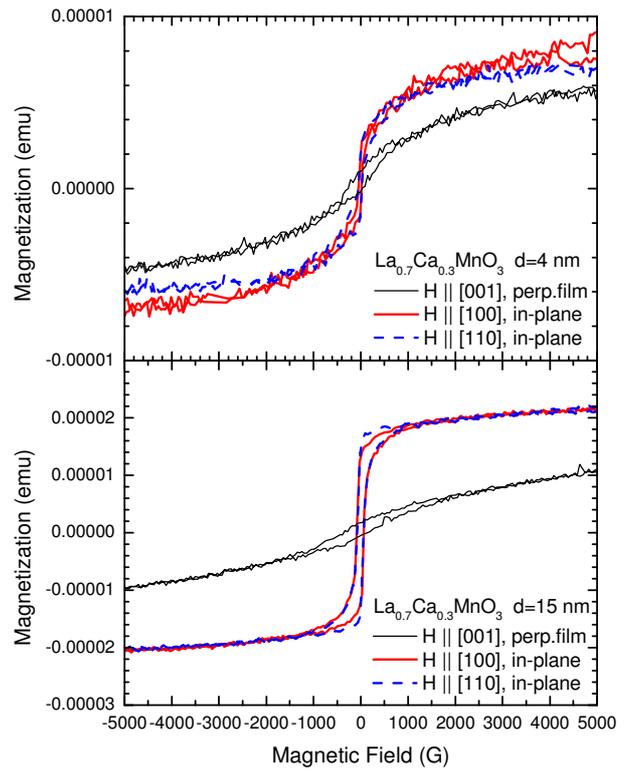

Fig. 1 (a) Magnetization hysteresis loops, (b) temperature dependent magnetisation (c) and resistivity of LCMO thin films with various thicknesses between 3 and 240 nm.

Fig. 2 Magnetization hysteresis loops of d=4 nm (*top*) and d=15 nm (*bottom*) LCMO thin films with the external magnetic field lying along [100] and [110] in-plane directions and along [001], perpendicular to the film-plane, at 77 K.

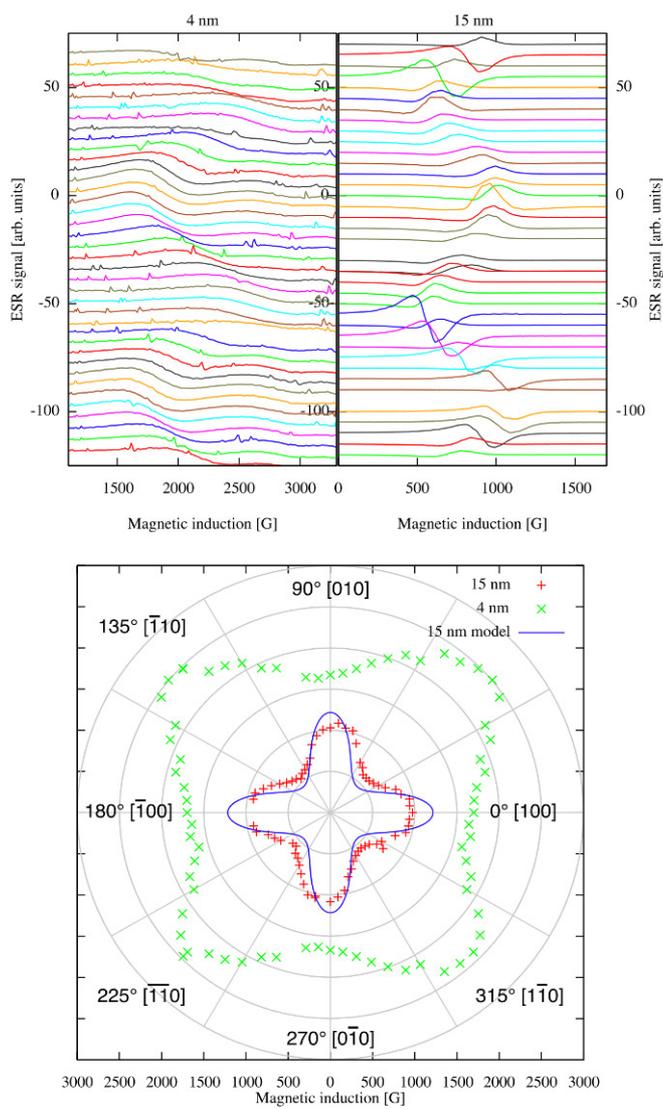

Fig. 3 (*top*) Ferromagnetic resonance spectra at f=8.88 GHz in d=4 nm (*left*) and d=15 nm (*right*) LCMO thin films at 77 K with the external magnetic field lying along various in-plane directions. The spectra were linearly displaced so that their baseline position indicates the in-plane angle. (*bottom*) FMR positions shown in polar-plot for both 4 and 15 nm thin films. The blue curve is a model for the 15 nm sample as described in the text. The radial axis corresponds to the magnetic field from 0 to 3000 G. Some of the 4 nm points are not independent data but mirrored from equivalent positions.

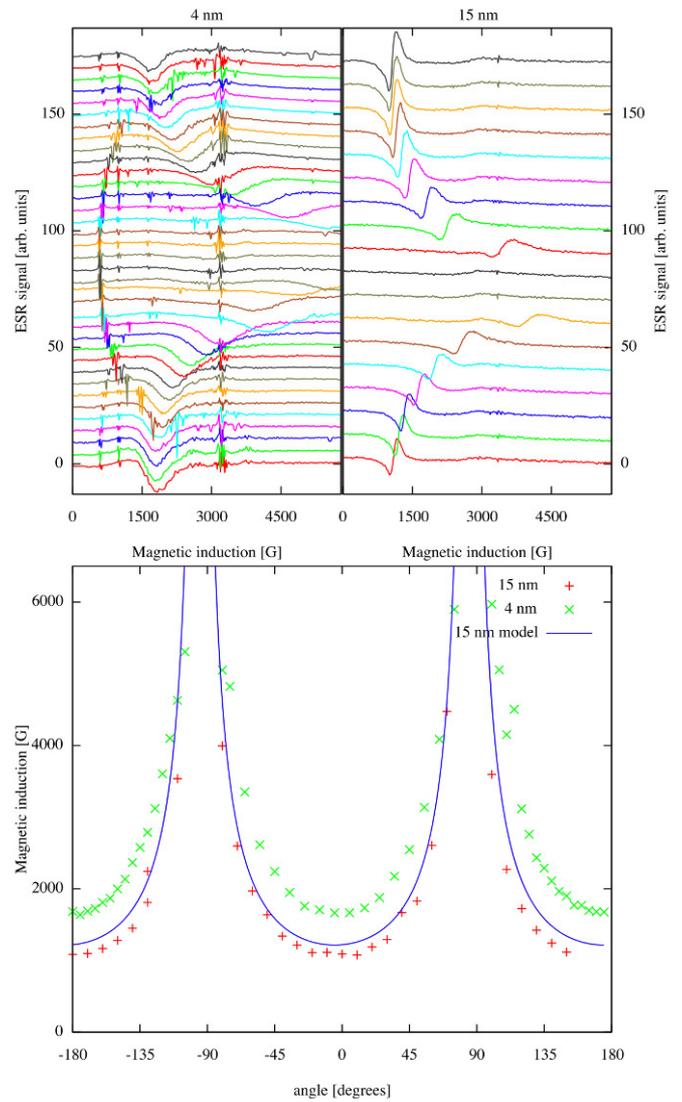

Fig. 4 (*top*) Ferromagnetic resonance spectra at f=8.88 GHz in d=4 nm (*left*) and d=15 nm (*right*) LCMO thin films with the external magnetic field rotated out-of-plane. The spectra were linearly displaced so that their baseline position indicates the out-of-plane angle, 90 being the field perpendicular to plane configuration. (*bottom*) FMR positions for both 4 and 15 nm thin films. The blue curve is a model for the 15 nm sample as described in the text.